\documentclass[12pt]{iopart}
\usepackage{iopams,mathrsfs}  
\usepackage[dvips]{graphicx}

\def\lb({\left(}

\def\hbar{\hspace{0pt}\raisebox{1pt}{$-$} \hspace{-7pt} h}

\newcommand{\eq}[1]{\begin{equation} #1 \end{equation}}
\newcommand{\eqs}[1]{\begin{eqnarray} #1 \end{eqnarray}}
\def\a{\alpha}

\def\g{\gamma}
\def\d{\delta}

\def\ee{\eta}
\def\th{\theta}

\def\m{\mu}

\def\x{\xi}

\def\p{\pi}
\def\r{\rho}
\def\s{\sigma}

\begin{document}

\title[Mass-degenerate Heavy Vector Mesons at Hadron Colliders]{Mass-degenerate Heavy Vector Mesons at Hadron Colliders}

\author{M. Piai and M. Round}

\address{Swansea University, School of Physical Sciences, Singleton Park, Swansea, Wales, UK}
\ead{461883@swan.ac.uk}
\begin{abstract}
We study the LHC phenomenology of two mass-degenerate heavy gauge bosons with the same quantum numbers as the $Z$ and $\gamma$.  We give a leading-order estimate of the number of events expected in Drell-Yan processes in  terms of the parameters of the model (mass and coupling) and of the LHC machine specifications (integrated luminosity and energy).  We consider the feasibility of measuring a forward-backward asymmetry  for various choices of the parameters and  estimate the potential reach.   We comment on how the results may affect future collider design and apply our results to a specific model of electro-weak symmetry breaking by way of example.
\end{abstract}

\section{Introduction}
The main purpose of the LHC is to discover the origin of electro-weak (EW) symmetry breaking and any associated new physics at the TeV scale.

One clean signal for the LHC would be the discovery of new, higher-mass copies of the standard-model (SM) gauge fields.  For such a scenario, the LHC reach in  observable masses and couplings is of interest. There are theoretical motivations for such searches.  Many weakly-coupled extensions of the SM gauge group yield new heavy vector fields~\cite{Zprime}. Such new fields also appear in little Higgs models~\cite{LH}.  Even  strongly-coupled extensions, such as  technicolor (TC), predict the existence of  towers of heavy spin-1 fields~\cite{TC}.

In TC models EW symmetry is broken by a condensate, along the lines of chiral symmetry breaking in QCD. 
Thus these models do not contain a light elementary Higgs particle~\footnote{In some strongly coupled models, with approximate conformal symmetry, a light dilaton might be present with interactions very similar to the Higgs of the SM~\cite{dilaton}. We do not consider this possibility.}.   As a result the most promising  test of these models at the LHC is a search for higher-mass spin-1 resonances.  In particular,  a copy of the full $SU(2)\times U(1)$ realized as heavy gauge bosons would be distinctive.   If the new bosons are nearly degenerate in mass then  the model will satisfy the precision EW constraints~\cite{PT, Barbieri}, without suppressing the coupling to SM fermions.  This situation would imply that the new boson resonances overlap at the LHC.

Once the LHC reaches its design specification, proton-proton collisions will scan over the energy range current data allow higher resonances to occur at.  In generic models this is in the  $1~\rm{TeV}$ region, but in many concrete realizations  the bound is above the $2-3~\rm{TeV}$ range~\cite{AdSTC, RS}.

In this paper, we consider the scenario in which a complete set of four spin-1 copies of the SM gauge bosons with heavy degenerate masses are produced in $pp$ collisions.  We concentrate on  detection of the neutral heavy-$Z$ and heavy-$\g$ through their decay into muons. The phenomenology of the degenerate scenario has been largely ignored to date in the literature.  This is in contrast to the single $Z'$ situation, see for example~\cite{ATLASTDR} and references therein.  For simplicity, we model the decay width of these particles by assuming it to be dominated by the decay into SM fermions.  This  avoids complications (and model-dependencies) arising when including the decay into light scalars and gauge bosons.   If the underlying dynamics is based on an $SU(N_c)$ gauge theory with large values of $N_c$ then the boson to boson decays are suppressed by the $N_c$ counting.

At LHC energies the SM background is mostly hadronic, so by looking at the neutral sector of higher resonances we exploit the very low background for muon and electron pair production.   Within these assumptions we can study the event rate to leading order.  Our aim is to identify what portion of parameter space will be testable at the LHC  as a function of integrated luminosity and energy.

In the SM, a forward-backward asymmetry arises from interference terms between 
$Z$ and $\g$ exchange diagrams.  It is a sizable effect, even in the TeV energy range.
Measuring such an effect at a hadron collider is onerous but a method and its use at the LHC was surveyed for instance by Dittmar~\cite{Dittmar}.  It was shown that a deviation of the measured asymmetry from the SM prediction is a possible signature of the presence of a new, very broad resonance with the quantum numbers of a $Z^{\prime}$.

In contrast we will study a relativity narrow resonance.  In our case, near the pole of some new resonance well into the TeV region the number of events is  well in excess of the SM.   So much so that the study of the di-lepton invariant-mass distribution  will contain a clear peak.  This will be a suitable discovery signature.    Furthermore for all practical purposes the SM contribution to this peak can be ignored.   In this case, one expects no significant asymmetry to be measurable at the peak from the SM.
 
If such a narrow peak is discovered in the data,  but it results from  the existence of a $\g'$, in addition to the $Z'$, then the interference between these two new states yields  a large asymmetry. Measuring this asymmetry  is hence a clean way to identify  the presence of two distinct particles  in the peak region, with masses so close together that they cannot be resolved in the invariant-mass distributions~\cite{deCurtis}.

The organization of our paper is the following.  We begin by introducing a model that encompasses the phenomenology of many specific models of EWSB.  We then summarize some known results about decay widths and cross-sections.   Then we proceed to study the Drell-Yan production of spin-1 resonances at the LHC.  To illustrate the dependence of the event rate on model parameters we make a series of approximations to give a simple formula for the total number of events near a peak.  We provide estimates of the discovery reach for two collider energies, 10 and 24 TeV.  

We then derive a simple formula for computing the reach of forward-backward asymmetry measurement at a collider.  At a 10 TeV machine the ability to search for a forward-backward asymmetry, of the type we are interested in, is limited to already excluded parameter space.  Therefore we present data at 14 TeV (which is close to the border of feasibility) and 24 TeV.  To illustrate the results  we finish by concentrating on a particular model.   Individual models give relations between the free parameters in our model-independent analysis and restrict the parameter space.  This allows us to evaluate a typical reach scenario.

A few words of explanation about these choices are needed.  The 10 TeV centre of mass energy analysis is representative of the LHC experimental program.  As we will see, large integrated luminosities are needed for discovery and hence comparing to the 7 TeV -- with low luminosity --  initial phase of LHC running is not very useful for the cases we study.  By 10 TeV energy, however, the reach is quite substantial.  This is the reason for our conservative choice of energy. The 24 TeV energy analysis is somewhat speculative, but provides a way to gauge how an energy upgrade compares to a luminosity upgrade of the apparatus, in reference to the specific process we study.

\section{Set-up and Rationale}
This section reviews cross-section and asymmetry calculations as well as defining the notations used throughout the paper. 
Most of the material in this section is well-known, but it is useful to remind the reader about it, and
to fix the notation used in later sections.
We model the new neutral resonances  
by adding an extra set of $SU(2)\times U(1)$ gauge fields  into the SM electro-weak Lagrangian.  
We assume that some mechanism spontaneously breaks
the $SU(2)^2\times U(1)^2$ gauge symmetry to the  SM $SU(2)_L\times U(1)_Y$ above the TeV scale,
and that in the process a copy of the $W$, $Z$ and $\gamma$ acquire a common mass $m$.  
In general, the latter is not rigorously correct:
the new heavy  bosons  would acquire different masses 
in any realization. However, the idea that the mass splitting is parametrically small is suggested by the precision electro-weak parameters.
We will provide a simple illustration of this in the next sub-section.
To parametrize the couplings of the new heavy gauge bosons to the SM currents we introduce a parameter $R$.     
This is defined so that  if the coupling of a given SM boson to a given SM current is $g_{SM}$ then the corresponding new heavy boson has a coupling of $\sqrt{R} g_{SM}$ to the same current. For example if the SM $Z$ coupling to an electron current is $g_Z$ then the corresponding new boson, the $Z'$, would have a coupling of $\sqrt{R} g_Z$ to the electron current.  
The common mass $m$ and the universal ratio $R$ of the couplings are the only new parameters in the model.  

It must be noted that in the paper we focus our attention on the neutral vectors only, setting aside the charged $W^{\prime}$.
This choice is due to the fact that for the processes we consider, the kinematics of the leptonic final state can be reconstructed 
precisely when the intermediate state is neutral, while this is difficult for charged states, in which case one has either  to
deal with missing energy, or to consider final states with hadronic jets.

\subsection{Mass Degeneracy}
As previously noted, in this work we  assume that the new heavy gauge fields are degenerate in mass.  As an example, illustrating the rationale behind this 
assumption, we examine the precision parameter $\hat{S}$~\cite{PT,Barbieri}.  Recall that on general grounds
one can write the precision  parameter $\hat{S}$ as
\eq{
\hat{S} =\frac{1}{4} g^2 \sum \left( \frac{f^2_v}{m^2_v}-\frac{f_{a}^2}{m_{a}^2}\right),
}with $g$ the $SU(2)_L$ coupling in the SM, $f$ and $m$ being the decay constant and mass of the vector ($v$) and axial ($a$) mesons. 
In models where  vector dominance holds the lightest among the states dominate the sum.  Denoting them states as the techni-$\r$ meson and the techni-$a_1$ meson we truncate the sum to these two states. 

Under this truncation, we write the parameters of these two states in terms of  a constant $a_1$ and a mass splitting $\Delta m^2$ ,
\eqs{
f_{a_1}^2 &=& f^2_\r + a_1 \frac{f_\r^2}{m_\r^2}\Delta m^2 ,\\
m_{a_1}^2 &=& m_\r^2+\Delta m^2.
}Working with  these definitions to leading order in $\Delta m^2/m_\r^2$ we see that,
\eq{\label{Shat}
\hat{S} =\frac{1}{4}g^2 \frac{f_\r^2}{m_\r^2}\frac{\Delta m^2}{m_\r^2} \left(1-a_1\right).
}
Experimentally,  $\hat{S}$ is found to be $O(10^{-3})$~\cite{Barbieri}.  This requires to implement some mechanism 
to obtain an acceptable value from eq.(\ref{Shat}).
Choosing $a_1=1$ would require some sort of fine-tuning, hence we assume that $a_1$ be some generic $O(1)$ number.
Another possibility, that   $f^2_\r/m_\r^2 \ll 1$,  is realized in QCD. However, this ratio scales as $N_c$ in the large-$N_c$ limit (for $N_c$ the number of colors of the confining gauge theory).  This makes it an unsuitable mechanism in a large-$N_c$ theory.
Also, allowing for  $f^2_\r/m_\r^2 \ll 1$ to be present would mean that the (techni-)rho mesons coupling to the fermions is heavily suppressed
(the parameter $R$ is very small),
and its main decay is into longitudinally polarized standard-model vector mesons. This is in analogy to what happens in QCD  where the rho meson decays into pions.
Finally, in this case the rho mesons are actually strongly coupled, in the sense that their width is large.  
For all of  these reasons we prefer not to use this suppression mechanism.  Another possibility to satisfy the electro-weak precision constraints is to assume that $\Delta m^2/m_{\r}^2\ll 1$, 
small enough that $\hat{S}\lesssim 10^{-3}$.  Interestingly, this observation extends beyond the vector-dominance approximation, at least at large-$N_c$, in the sense that $\hat{S}$ is suppressed if the mass splitting between new axial and vector meson states is small.  This is the regime which inspires this present work.

\subsection{Parton-level processes}
 In this section we begin the computation of the relevant cross-section, following standard textbook procedures.  Initially this is done for a lepton machine then the results are extended to a hadron machine.  This choice is made to highlight the physics of the model considered. 
 
 We find it useful to introduce the function $P_V$, 
\eq{
P_V=(\hat{s}-m_V^2+i m_V \Gamma_V)^{-1},
} where $\sqrt{\hat{s}}$ is the total energy, in the centre of mass frame, for the collision whose cross-section we compute, $m_V$ is the mass of the  particle $V$ and $\Gamma_V$ is the decay width of the particle $V$ for $V=\g, Z,\g^{\prime},Z^{\prime}$.  This function enters  the propagator
of the boson $V$.  Note that this requires computing the decay widths for the two new resonances.  The new heavy gauge bosons can also decay into SM gauge boson pairs (e.g $Z' \rightarrow W^+ W^-$). However  the rate of such processes is very model-dependent, and at the same time assuming it to dominate the decay width might imply that the new gauge bosons be strongly coupled, as anticipated.  
In this study we assume such channels are negligible.  This is equivalent to setting the 3-point and 4-point meson vertices to zero.   This assumption is justified by the $1/N_c$ counting if the underlying dynamics is similar to that of  a large-$N_c$ gauge theory and the 
heavy spin-1 states are mesons of the strongly coupled sector, in analogy with  the $\rho$ and $a_1$ of a (large-$N_c$) QCD-like theory.

Within these assumptions, the $\g'$ and $Z'$ have decay widths proportional to those of the $\g$ and $Z$ in the SM. The constant of proportionality is the ratio between the SM and new gauge boson couplings.  This is the parameter $R$ introduced earlier on.  
Denoting the $SU(2)$ coupling of the standard model by $g$,  the Weinberg angle by $\th_w$, 
the fine structure constant by $\a$ (and the new heavy boson mass by $m$) the decay widths into SM fermions of the new states  are,
\eqs{\Gamma_{\g'} &=& \frac{8}{3}R\a  m,\\
\Gamma_{Z'} &=& \frac{R g^2m }{4\p \cos^2 \th_w }\left(1-2\sin^2 \th_w +\frac{8}{3}\sin^4 \th_w \right)\,,
}
where we summed over all the SM fermions, including the top, neglecting the fermion masses.  Typical values of the new heavy  masses are $O(1)$ TeV, thus the error introduced by neglecting fermion masses in the phase-space factors is negligible.

As an illustration of the model, we begin by removing the complications of $pp$ collisions and take the case of $e^+e^- \rightarrow \m^+\m^-$.  There are four possible diagrams.  In addition to the SM channels of $\g$ and $Z$ exchange there is also $\g'$ and $Z'$ exchange.  Following standard notation, introduce $c_V$ and $c_A$ for the vector and axial couplings defined as
\eqs{
c_V&\equiv &  T^3 -2 Q \sin^2 \th_w ,\\
c_A&\equiv & T^3,
}where $Q$ denotes the electric charge and $T^3$ the third component of weak isospin for the relevant fermion.  We denote
\eqs{
\Pi_\g (\hat{s}) &=& 4\p \a Q_i Q_f (P_\g+RP_{\g'}),\\
\Pi_Z (\hat{s}) &=& \frac{g^2}{4\cos^2\th_w} (P_Z+RP_{Z'})\,.
}for $Q_i$ and $Q_f$ the charges of the initial and final fermions in the Feynman diagram.  
Using this notation the angular $2 \rightarrow 2$ partonic cross-section  can be written out as,
\eq{
\frac{d^2\hat{\s}}{d\hat{\Omega}}= \frac{\hat{s}}{64\p^2} \left\{A(1+\cos^2 \hat{\th}) + B\cos \hat{\th}\right\},
} for $\hat{\th}$ the angle of inclination with respect to the beam and $\hat{\Omega}$ the solid angle of the scattering process.  The $A$ and $B$ are defined as,
\eqs{
A &=& |\Pi_\g |^2 + 2c^2_V\Re (\Pi_\g \Pi_Z^*)+ (c_V^{2}+c_A^{2})^2|\Pi_Z |^2\\
B &=& 4 c_A^2\Re (\Pi_\g \Pi_Z^*) +8c_V^{2}c_A^{2} |\Pi_Z |^2
}where $\Re$ denotes the real part.

Similarly we can write the forward-backward asymmetry $A_{FB}$ as 
\eq{
A_{FB}=\frac{\left(\int_0^1-\int^0_{-1}\right) d\cos\hat{\th} \, d\hat{\s}/d\cos\hat{\th}}{\left(\int_0^1+\int^0_{-1}\right) d\cos\hat{\th} \, d\hat{\s}/d\cos\hat{\th}}.
}  We obtain,
\eq{
A_{FB}=\frac{3B}{8A}.
}

It is useful to combine the above expressions and rewrite the cross-section in terms of an angular function $g(\hat{\th})$:
\eq{
\frac{d^2\hat{\s}}{d\hat{\Omega}} = \frac{8 \hat{s}A}{192 \p^2} g(\hat{\th})\,,
}
with
\eq{\label{Eq:ghat}
g(\hat{\th})=
\frac{3}{8}(1+\cos^2 \hat{\th})+A_{FB}\cos \hat{\th}\,.
}This form of the partonic cross-section, in terms of $g(\hat{\th})$, is the most useful form for our discussion of asymmetry.

Above an energy of 500 GeV the SM  contribution becomes negligible for our purposes,
as anticipated in the introduction.  Events  near the resonance of the new heavy states, $\hat{s}\simeq m^2$, are dominated by the new states  as long as  $m\Gamma_{V'} \ll \hat{s}$, since in this case
\eqs{
|P_V|&\simeq& \hat{s}^{-1},\\
|P_{V'}| &\simeq & (m\Gamma_{V'})^{-1}.
}
Therefore unless $R$ is very small the new resonances will dominate.  
 The definitions of $\Pi_V$ given can now be approximated by dropping the $P_V$ terms.  
\eq{
\Pi_V \propto P_{V'}.
}
 We use this approximation to re-write the asymmetry near the new resonance ($\hat{s}\simeq m^2$) as
\eqs{
A_{FB}& =& \frac{3}{2}\frac{c_A^2\Gamma_{\g'} \Gamma_{Z'} + 2 c_V^2c_A^2\Gamma_{\g'}^2}{\Gamma_{Z'}^2
+ 2c_V^2\Gamma_{\g'}\Gamma_{Z'} + (c_V^2+c_A^2)^2\Gamma_{\g'}^2}.
}This approximation is used in computing the numerical values of $A_{FB}$ that we work with later.  This method avoids the complications introduced by accounting for the weak dependence of $A_{FB}$ on $\hat{s}$ .

Now we quote the correct formulae in the case of a hadron collider.   The formulae for asymmetry and cross-section at partonic level can be amended by adding a factor for the number of colors $N_c$, and changing to quark couplings.  Denoting the quark/lepton couplings with a $q$/$l$ superscript, $A$ is modified to
\eqs{\label{Eq:amplitude}
A_{q\rightarrow l} &=& |\Pi_\g |^2 + 2c^q_Vc_V^l\Re (\Pi_\g \Pi_Z^*)\\
\nonumber&&+({c_V^q}^{2}+{c_A^q}^{2})({c_V^l}^{2}+{c_A^l}^{2})|\Pi_Z |^2.
} Including the appropriate color factor, the partonic cross-section for $q\bar{q}\rightarrow l^+l^-$ is
\eqs{\label{Eq:cs}
\hat{\s}&=& \frac{1}{N_c} \frac{\hat{s}}{12\p} A_{q\rightarrow l},
}where the angular dependence has been integrated over.  
Adjusting $B$ to account for quarks gives
\eq{
B_{q\rightarrow l} =  4c^q_Ac_A^l\Re (\Pi_\g \Pi_Z^*) +8c^q_Vc_V^lc^q_Ac_A^l |\Pi_Z |^2.
}The asymmetry  at the partonic level is hence
\eq{\label{Eq:as}
A_{FB} = \frac{3B_{q\rightarrow l}}{8A_{q\rightarrow l}}.
}

In summary, our study of event rates  focuses around the partonic cross-section eq.(\ref{Eq:cs}).  Our asymmetry calculations makes use of the angular partonic cross-section
\eq{
\frac{d^2\hat{\s}}{d\hat{\Omega}} = \frac{8 \hat{s}A_{q\rightarrow l}}{192 N_c \p^2} g(\hat{\th})
}The forward-backward asymmetry required in $g(\hat{\th})$ is computed from eq.(\ref{Eq:as}) under the approximation of no SM background.  Explicitly this means that
\eqs{
A_{FB}& =& \frac{3}{2}\frac{c_A^q c_A^l\Gamma_{\g'} \Gamma_{Z'} + 2 c_V^a c_V^l c_A^q c_A^l\Gamma_{\g'}^2}{\Gamma_{Z'}^2
+ 2c_V^a c_V^l\Gamma_{\g'}\Gamma_{Z'} + ({c_V^q}^{2}+{c_A^q}^{2})({c_V^l}^{2}+{c_A^l}^{2})\Gamma_{\g'}^2}.
}
 
\subsection{Drell-Yan processes}

We will follow the standard recipe to model processes at a hadron collider.  We  consider the incident protons to be composed of partons holding fraction $\x$ of the total proton momentum $p$.  Numerous sets of parton distribution functions (PDF) are available through Fortran code~\cite{lhapdf}.  For this paper we choose the Fermi02 PDF set~\cite{Fermi}.  For practical purposes, one can treat the proton and its constituent
 partons as massless.  The two beams of protons each have energy $\sqrt{s}/2$.  
The energy of a parton-parton collision $\hat{s}$ is given, in terms of the two fractions $\x_1$ and $\x_2$, by
\eq{
\hat{s}=\x_1\x_2 s.\label{condition}}  If, for a parton $x$, the PDF is denoted by $f_x(\x)$ we can follow convention and write the $pp \rightarrow l^+l^-$ cross-section as,
\eq{
\frac{d^2\s}{d\x_1 d\x_2 } =\sum_q [ f_q(\x_1) f_{\bar{q}}(\x_2) + f_q(\x_2) f_{\bar{q}}(\x_1)]\hat{\s}.
}
To fix the total energy $\sqrt{s}$ of the $pp$ collision,  we impose the condition of Eq.~(\ref{condition}).  This means we may define the parton luminosity $\mathscr{L}_q$ to be
\eqs{
\mathscr{L}_q &\equiv& \int_0^1 \d(\hat{s}-\x_1\x_2s)  [f_q(\x_1) f_{\bar{q}}(\x_2)+\x_1\leftrightarrow\x_2] d\x_1 d\x_2 , \nonumber\\
&=&\frac{2}{s} \int_{\hat{s}/s}^1 \frac{1}{\x} f_q(\x) f_{\bar{q}} \left(\frac{\hat{s}}{s\x}\right)d\x
}and write the full cross-section summed over all quarks as,
\eqs{
\frac{d\s}{d\hat{s}}=\sum_q \hat{\s} \mathscr{L}_q.
}Finally we can write the total number of events $N$ in some range of $\hat{s}$ at a machine with integrated luminosity $L$,
\eq{
N=L\int \frac{d\s}{d\hat{s}} d\hat{s}.
}

We allow $\sqrt{s}$ to vary between 10 and 28 TeV.  The coupling ratio $R$ will be allowed to vary between 0 and 1, where the gauge boson coupling is certainly perturbative.  Then the mass range will be taken as lying between 1 and 4 TeV.  Analysis will also require the assumption that decay widths are small compared to the resonance mass.  Inspection of the decay widths, under the limits on $R$ and $m$, shows this condition is satisfied.

Asymmetry measurements at a hadron collider are made more difficult (in comparison with a leptonic machine) by the unknown directions of the initial  quark and anti-quark.  
However, at high rapidity there is a much greater chance that the quark 
carries the bulk of the momentum.   
Thus if data near a resonance in the invariant-mass distribution 
are cut so that only events with a rapidity
$Y$ greater than some  value $Y_c$ ($|Y|>Y_c$) are used,
 then there is 
a corresponding degree of certainty regarding the quark directions.  
This method is used and explained more fully by Dittmar~\cite{Dittmar}. 
 In summary,  the magnitude of the rapidity controls the probability of misidentification of the quark/anti-quark directions, while the sign of the rapidity represents whether the quark was from the left or right in the collider.  This methodology of inserting cuts,  Lorentz boosting with the rapidity modulus to the muon centre of mass frame and finally using the sign of $Y$ to define the angle consistently allows one to infer an asymmetry.  Such an asymmetry is the remnant of the partonic asymmetry.

\section{Model Independent Study}
\subsection{Event Rate}

In this section we introduce a set of approximations that are
useful in order to study the dependence of observable quantities
(such as number of expected events and forward-backward asymmetry)
on the parameters of the model and on the machine specifications.
The numerical part of the analysis will, however, use the complete 
equations and PDF distributions without approximations.  From the approximations made we gain insight into how event rates behave as a function of model parameters.

Using standard electro-weak relations the coupling $g$ can be written 
in terms of $\a$ and $\sin^2\th_w \simeq 0.22$.  Hence, near the new resonance:
\eq{
\frac{\Pi_{\g}}{\Pi_{Z}}\simeq -\frac{3}{2} Q_q \left(1-2\sin^2\th_w+\frac{8}{3}\sin^4\th_w\right)\,\simeq\,-Q_q\,.
}
Using Eqs.~(\ref{Eq:amplitude})-(\ref{Eq:cs})
we can  comment on the relative strength of the $\g'$ and $Z'$
contributions to the total cross-section. Because $c_v^l  \ll 1$,
the cross term in Eq.~(\ref{Eq:cs}) is suppressed. Also, in the last term of this equation
$c_V^{f\,2}\ll c_A^{f\,2} \simeq 1/4$, and hence the contribution of the $\g^{\prime}$
is a factor of $\sim 16 Q_q^2$ larger than the contribution coming from the $Z^{\prime}$.
All of this, together with the fact that the PDF for the up-quark and down-quarks are the most important, 
and that the coupling of the up quarks to $\g'$ is four times stronger 
than the coupling of down quarks, means that the most significant contribution to the total event  rate 
is given by the subprocess $u\bar{u} \rightarrow \g' \rightarrow \m^+\m^-$.  The next largest contribution to the event rate is from the cross term between the new photon and $Z$-like states (which is approximately half the value of the leading contribution) and is responsible for the asymmetry we discuss later.

Focusing on the $u$-quark PDF, we may attempt to approximate it with some form of expansion.  The only  relevant dimensionless combinations are $\hat{s}/s$ and $m^2/s$.  As we are looking at the peak we can replace the square mass with $\hat{s}$ so that only one relevant dimensionless combination remains.  From the definition of $\mathscr{L}$ the overall dimension is $1/s$.  If the mass of the resonance were to increase one would expect the event rate to decrease.    It is also reasonable to think increasing the machine energy $s$ will increase the event rate.  We use an approximate form of,
\eq{
\mathscr{L}_u \sim \frac{1}{s}\left(\frac{\hat{s}}{s}\right)^n,
}and consider various $n$ so as to obey the above criteria and then numerically fit the constant of proportionality.  The minimal $n$ would be $-2$ to fulfill our expectations for the behavior of $\mathscr{L}$.  Viewing our expression as the leading term in some expansion a polynomial can be added to increase accuracy.  We find that an acceptable approximation is,
\eq{\label{Eq:approxPDF}
\mathscr{L}_u  \sim \frac{s}{\hat{s}^2} \cdot \frac{1}{1+a_1 (\hat{s}/s)^2 +a_2 (\hat{s}/s)^4 +\ldots}\,,
}
provided $\sqrt{\hat{s}} \in [1,4] \,{\rm TeV}$. 
We then go through a numerical fit of the constants $a_1$ and $a_2$. In the range $\sqrt{s} \in [10,28]~\rm{TeV}$ just those two terms are needed.

We are interested in the event rate $N$ in the peak region
\eqs{
N&\propto &\int \sum_q \frac{d\s}{d\hat{s}} d\hat{s},\nonumber\\
&\simeq & \int \mathscr{L}_u \hat{\s}_u d\hat{s}.
}
Approximating ${\mathscr L}_u$ with its value at the pole (which is allowed since the 
resonances are very narrow), and using
\eqs{
\int \frac{\hat{s}}{(\hat{s}-m^2)^2+m^2\Gamma_{\g'}^2}d\hat{s} \simeq 2\p \frac{m}{\Gamma_{\g'}}.
}
yields
\eqs{
 \int \mathscr{L}_u \frac{d\hat{\s}_u}{d\hat{s}}d\hat{s} &\simeq& \mathscr{L}_u(m^2) R^2 2\p \frac{m}{\Gamma_{\g'}}\\
&\propto &  R  \mathscr{L}_u (m^2)\,,
}
where  the SM background from ${\g}$ and $Z$ has been neglected.

Finally,  the contribution  of up-quarks to the total rate is,
\eq{
N_u\propto \frac{sLR}{m^4}\cdot  \frac{1}{1+a_1 m^4/s^2 +a_2 m^8/s^4}\,,
}
up to a proportionality constant that contains the electro-weak couplings and 
the normalization of the PDF distribution.

To correct for the neglected contributions of $d\bar{d}$ events and $Z'$ exchange we can consider the ratio of the found rate $N_u$ to a more complete expression involving $u\bar{u}$ and $d\bar{d}$ events with $\g'$ and $Z'$ exchange.  For notational simplicity let us absorb the coupling coefficients into the definition of $P_V$ briefly and write the ratio $\eta$ as,
\eq{
\frac{1}{\eta}\equiv\frac{\mathscr{L}_u\int\hat{s}|P_{\g'}|^2d\hat{s}}{ \mathscr{L}_u\int\hat{s}|P_{\g'}+P_{Z'}|^2d\hat{s}+\mathscr{L}_d\int\hat{s}|P_{\g'}+P_{Z'}|^2d\hat{s}}.
}

An expression similar to 
Eq.~(\ref{Eq:approxPDF}) describes, with different  choices of the coefficients,
${\mathscr  L}_d$.
 Then we find that a very good approximation is given by
\eq{\frac{1}{\eta}\,=\,\label{Eq:eta0}
\frac{1+b_1(\hat{s}/s)^2 +b_2 (\hat{s}/s)^4}{1+a_1(\hat{s}/s)^2 +a_2 (\hat{s}/s)^4}(1+ c_1 m + c_2 R)\
}for some $b_i$ and $c_i$ that are combinations of the up and down coefficients.  

The total event rate can be written, up to an overall constant, as
\eq{\label{Eq:eta}N\simeq N_u\ee.}
We fix the parameters, $a_1,a_2,b_1,b_2,c_1,c_2$ plus an overall normalization with the
 direct numerical computation using the PDFs.  
 We estimate the  error  hence introduced between the estimated event rate and the direct computation to be about $5\%$.

\begin{figure}
\centering
\includegraphics{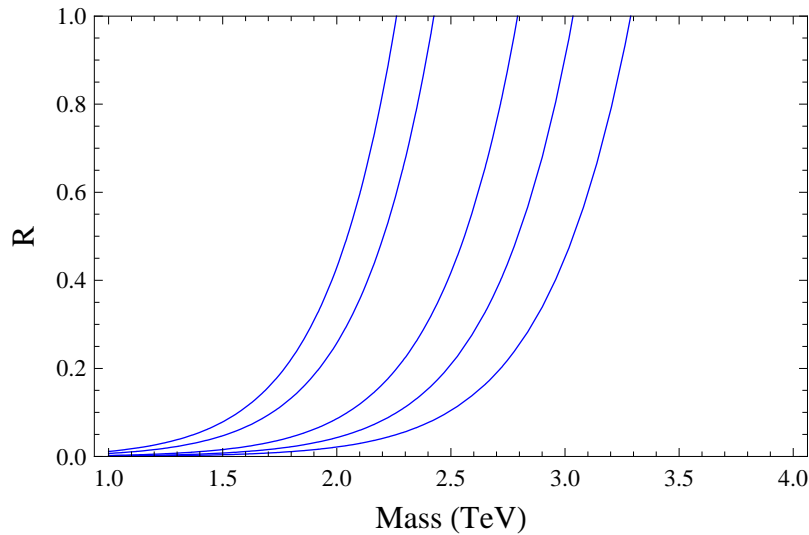}
\caption{\label{Event10}The ratio $R$ between the couplings of the new and SM boson to SM currents, such that the number of events $N$ divided by the integrated luminosity $L$ is fixed, and plotted as a function of the heavy boson mass at a collision energy of $\sqrt{s}=10$~TeV.  The leftmost line is for $N/L=5~\rm{fb}$ then reading from left to right the lines are, $N/L=3,1,1/2~\rm{fb}$, finishing on the rightmost line of $N/L=1/4~\rm{fb}$.  The events $N$ are for the production of muon pairs only.}
\end{figure}

\begin{figure}
\centering
\includegraphics{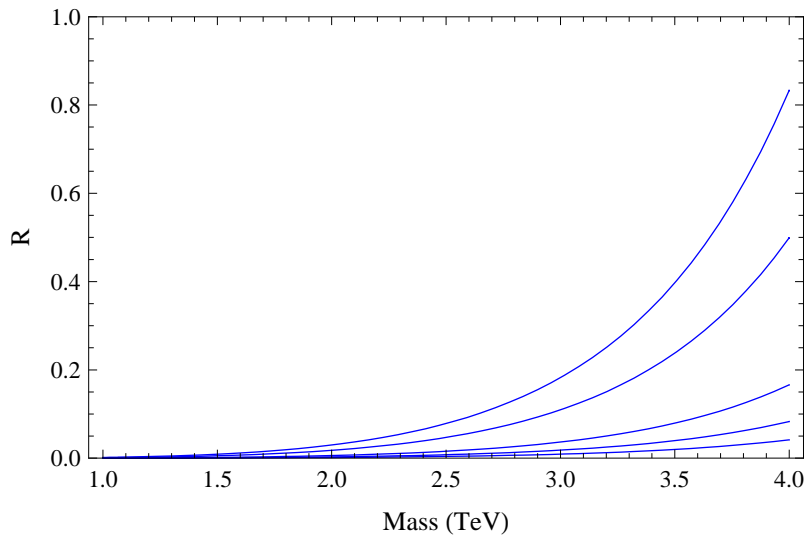}
\caption{\label{Event24}The ratio $R$ between the couplings of the new and SM boson to SM currents, such that the number of events $N$ divided by the integrated luminosity $L$ is fixed, and plotted as a function of the heavy boson mass at a collision energy of $\sqrt{s}=24$~TeV.  The uppermost line is for $N/L=5~\rm{fb}$ then reading from top to bottom the lines are, $N/L=3,1,1/2~\rm{fb}$, finishing on the lowest line of $N/L=1/4~\rm{fb}$.  The events $N$ are for the production of muon pairs only.}
\end{figure}

Discovery of the new resonance requires detecting a minimum number of events $N$ at a machine integrated luminosity $L$.  The approximate event rate can be solved for $R$, given $m$ at various machine energies and integrated luminosities,
hence producing Figs.~\ref{Event10} and~\ref{Event24}.  These two plots show 
how the experimental reach changes when improving the energy of a hadron collider, in comparison with 
increasing only its luminosity. We conclude this subsection with a couple of examples, to help the reader in  interpreting the plots.

Suppose one considers $N=10$ events to be necessary for discovery of a new resonance.  
Setting $N=10$, if one has data corresponding to a $40 ~\rm{fb}^{-1}$ integrated luminosity,
 then the line $N/L=1/4$ fb indicates the minimum coupling $R$ required to see such an effect.  
 In the case of a 10 TeV machine, for a new resonance with mass $m=3$ TeV, 
 then the minimum $R$ is around 0.4.  Therefore any new boson with coupling ratio $R>0.4$ and $m<3$ TeV will produce at least 10 events 
 in the data set obtained with $40~\rm{fb}^{-1}$ integrated luminosity.

Changing the machine energy from 10 TeV to 24 TeV means 
moving from Fig.~\ref{Event10} to Fig.~\ref{Event24}.  At $10$ TeV, by improving the integrated luminosity
by a factor of $20$, 
 one may move from a reach in masses below 2.2 TeV to roughly 3.3 TeV, provided $R=1$ (see the leftmost and rightmost lines in Fig.~\ref{Event10}).  
Whilst improving the collision energy takes the reach to well  above 4.0 TeV, even with the no improve in integrated luminosity, for the same $R=1$
(see the leftmost curve in Fig.~\ref{Event24}).

\subsection{Forward-Backward Asymmetry}
In the present study we limit ourselves to theoretical considerations only, so that we may ascertain how feasible asymmetry measurements are from a  theoretical viewpoint.  In fact one may argue that, due to large PDF uncertainties, a full scale  simulation of the experiments  
is overkill before LHC running. It would be interesting to see a detailed 
simulation once LHC data becomes available, and new data about the PDFs is available.

We restrict our attention to two main sources of 
error in the asymmetry extraction algorithm: the potential for misidentification of the quark direction and 
finite number of events passing the rapidity cut $Y>Y_c$.  
 We set aside experimental errors, coming for instance 
 from the measurement of the energy of the final leptons.  

 Consider the angular  distribution in Eq.~(\ref{Eq:ghat}) as a probability distribution function.  
This distribution has a mean defined as
\eq{
\langle \hat{\th}\rangle=\int\hat{\th}g(\hat{\th})\sin\hat{\th}d\hat{\th}=\frac{\p}{4}(2-A_{FB})\,,
}
and a variance defined by
\eq{
\sigma_{p}^2\equiv\int\left(\hat{\th}^2-\langle \hat{\th}\rangle^2\right)g(\hat{\th})\sin\hat{\th}d\hat{\th}\,.
}
Interestingly, $A_{FB}$ turns out to be  just proportional to $\langle \hat{\th}\rangle$. It is possible to estimate the deviation of the mean $\langle \hat{\th}\rangle$ from its true value as $\s_p/\sqrt{n-1}$ where $n$ is the number of events used to study the angular distribution.  Hence we estimate the difference between the true and measured asymmetry due to statistics to be,
\eq{
\Delta A_{FB}^{\rm stat} = \frac{4\s_{p}}{\p\sqrt{n-1}}\,.
}Here $n$ should be thought of as representing the number of events remaining after a cut in rapidity is applied to the total number of events $N$.

Secondly there is a probability of quark misidentification introduced by the methodology. 
The assumption that the initial quark direction be always the one 
of the out-going $\g^{\prime}$ or $Z^{\prime}$, implies 
a possible misidentification  $\hat{\th}\leftrightarrow \pi-\hat{\th}$.
 Let the probability of misidentification be $p(Y_c)$, then in a sample of $n$ events $np(Y_c)$ are
  misidentified.  Let $n_F$ ($n_B$) events travel in the forward (backward) direction in the collider.  Then $n=n_F+n_B$ and the forward-backward asymmetry is defined as $(n_F-n_B)/n$ in this notation.  The worst case scenario would be that all $np(Y_c)$ events were placed in $n_F$ but should have been placed in $n_B$ or vice versa.  In this case the difference in true and measured asymmetry would be,
\eqs{
 \Delta A_{FB}^{\rm sys}&=& \pm 2p(Y_c)
}Adding the misidentification  and the statistical together we can define a  `total misidentification' of,
\eq{
\Delta A_{FB}= \frac{4\s_{p}}{\p\sqrt{n-1}} +  2p(Y_c),
}We conservatively add the two  linearly.

Now we define $P(Y_c)$ to be the fraction of events  with $|Y|>Y_c$, i.e. the probability an event survives the chosen rapidity cut.  
Then we can relate the number of events in the peak, $N$, to the number used for the asymmetry extraction, $n$, through $P(Y_c)$ by,
\eq{
n=P(Y_c)N.
} Suppose we  are interested in measuring an asymmetry at some level of significance.  To do this introduce  $\mathscr{E}$ and require that $\Delta A_{FB} < \mathscr{E}$.   We must impose  $\mathscr{E}>2p(Y_c)$ so that the chosen cut value is capable of producing data of sufficiently low misidentification.  This  implies that we require,
\eq{
n>  \left(\frac{4\s_{p}}{\p(\mathscr{E}-2p(Y_c)}\right)^2+1
} or equivalently,
\eq{
N>\frac{1}{P(Y_c)}\left[\left(\frac{4\s_{p}}{\p(\mathscr{E}-2p(Y_c)}\right)^2+1\right]
}total events.  

For the model considered, at the partonic level the asymmetry is $A_{FB}\simeq -0.5$
in the case of $u\bar{u} \rightarrow \m^+\m^-$, while for $d\bar{d}$ 
events the asymmetry would be $A_{FB} \simeq-0.15$. (For  comparison,
SM exchange plus only one new resonance 
the asymmetry vanishes up to 3 decimal places.)
The measured asymmetry will then be some average of  the two.  
Using what we showed before, namely that the $u\bar{u}$ events give a
 larger contribution  to the total rate,  
we estimate the combined asymmetry to be
$A_{FB}\sim-0.42$.
Therefore seeing an asymmetry at a collider would indicate that there are two particles in 
the new neutral resonance.

To measure such an effect at a collider one would require a `total misidentification' $\mathscr{E} < 0.14$ in order to conclude, at the 3$\s$ level, an asymmetry is present.  Consider the case that all events happen to fall precisely at the rapidity cut $Y_c$.  Then the probability $p(Y_c)$ is calculable, yielding a conservative estimate:  with real data having $|Y|>Y_c$ one can in principle use a larger value of  $p(Y_c)$.  
The fraction $P(Y_c)$ can also be computed. 

Optimal choice of $Y_c$ for a minimal $N$ can be achieved by writing a simple minimization algorithm.  A high $Y_c$ gives high quality data (i.e. contaminated by small misidentification systematics)
with low $p(Y_c)$ but few events will pass such the cut, reducing also $P(Y_c)$.  
The converse will be true for a low cut $Y_c$.  
We used an optimization algorithm finding a $Y_c$ such that the required number of 
events has been minimized.  

\begin{figure}[hbp!]
\centering
\includegraphics{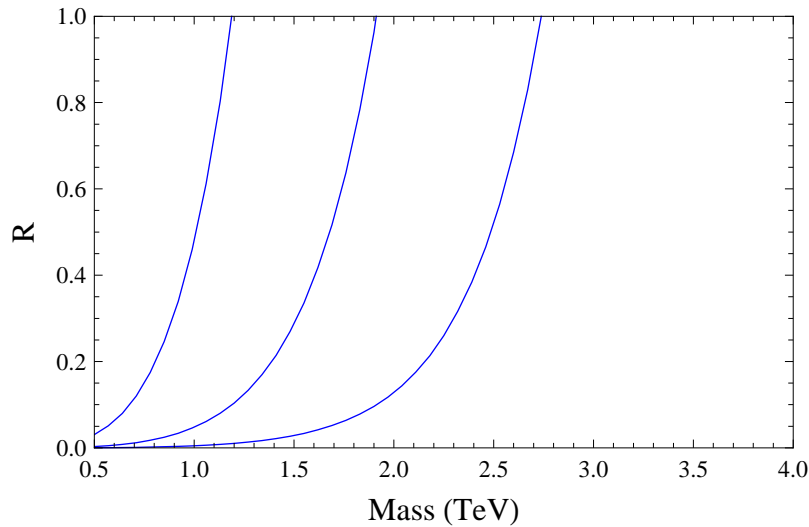}
\caption{\label{AFBmeasure14}The minimum ratio $R$ between couplings of the new and SM boson to SM currents, such that at the $3\s$ level (see text for definition) a forward-backward asymmetry of -0.42 is detectable, plotted for various integrated luminosities at a collision energy of $\sqrt{s}=$14 TeV and as a function of the heavy boson mass.  The leftmost line is for an integrated luminosity of $1~\rm{fb}^{-1}$, the middle line for $10~\rm{fb}^{-1}$ and the rightmost $100~\rm{fb}^{-1}$.}
\end{figure}

\begin{figure}[hbp!]
\centering
\includegraphics{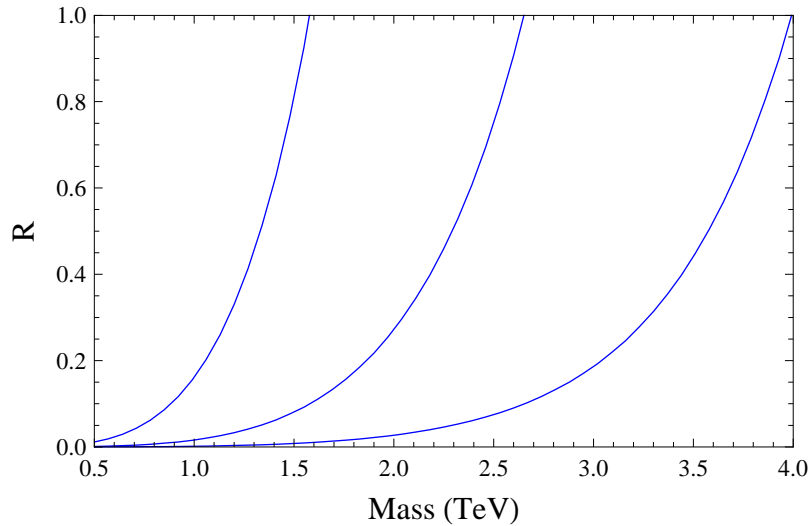}
\caption{\label{AFBmeasure24}The minimum ratio $R$ between couplings of the new and SM boson to SM currents, such that at the $3\s$ level (see text for definition) a forward-backward asymmetry of -0.42 is detectable, plotted for various integrated luminosities at a collision energy of $\sqrt{s}=$24 TeV and as a function of the heavy boson mass.  The leftmost line is for an integrated luminosity of $1~\rm{fb}^{-1}$, the middle line for $10~\rm{fb}^{-1}$ and the rightmost $100~\rm{fb}^{-1}$.}
\end{figure}

We compute the minimum number of events required for a $3\s$ uncertainty at given mass.  Using the approximate event rate given by eqs.~(\ref{Eq:eta0}) and~(\ref{Eq:eta}) we convert the number of events into a statement about $R$.  The results of this process are shown in Figs.~\ref{AFBmeasure14} and~\ref{AFBmeasure24}.
Figure~\ref{AFBmeasure14} shows that large increases in integrated luminosity lead to modest gains in reach,
compared to the gain obtained by going to larger energies, as shown in Fig.~\ref{AFBmeasure24}.  
One can also see from fig. \ref{AFBmeasure14} that in order to test the interesting range $m>2 ~\rm{TeV}$, one requires a very high integrated luminosity of $100 ~\rm{fb}^{-1}$.   At a 10 TeV machine it is not possible to measure an asymmetry of a neutral heavy gauge boson in this range of most interest,
and hence we omit showing the related plot.

\section{PDF Uncertainties}
So far the uncertainty on the PDFs themselves has been completely ignored.  The Fortran module supplied by~\cite{lhapdf} has the capability to compute PDF uncertainties.  As the computational cost to produce uncertainties with the package is very large, we use an approximate expression which captures the essential features of the PDF uncertainties.  Namely, the quark PDFs are well known up to fractions of around 0.5, then they become poorly understood while the anti-quark PDFs are poorly understood across all of the fraction range.  Two piecewise functions, one for the $u$ and one for the $\bar{u}$, approximate this behavior
\eqs{
\frac{\Delta u(\x)}{u(\x)} &=&  \cases{0.05& if $\x < 0.5$\\0.8\x -0.35&otherwise}\\
\frac{\Delta \bar{u}(\x)}{\bar{u}(\x)} &=& \cases{2.25\x &if  $\x < 0.8$\\25\x -18.2&otherwise}
}
These uncertainties are approximations to the results given by using the package~\cite{lhapdf} for our chosen PDF set~\cite{Fermi} and the results broadly agree with those published by the CTEQ collaboration for their PDF sets~\cite{cteqerrors}.  Even if  an event is not identified as being $u\bar{u}$ or $d\bar{d}$ in our analysis,  the $u\bar{u}$ dominates events and hence the uncertainty indicated will offer a good guide as to the situation.  

As the asymmetry extraction methodology involves cutting the data for large $Y$, 
it  probes the region with
  a smaller anti-quark fraction and a larger quark fraction.  
  If the quark fraction gets above 0.5 the uncertainty  grows rapidly.  
  At a 10 TeV collider this effect will come in faster as high fractions are required to make large invariant masses while the effect will be slower to appear in a 24 TeV collider.

In general, most events in a peak will have low rapidity.  As a result we 
expect far lower uncertainties for event rate measurements compared to asymmetry measurements.  

\section{An Example Model}

In order to illustrate the results of the model-independent study it is useful to apply them to a specific model.
 For example in the AdS/TC model in~\cite{Piai2}  a 5D model is formulated  in AdS space and bounds on the $(R,m)$ parameter space are discussed.  

\begin{figure}
\centering
\includegraphics{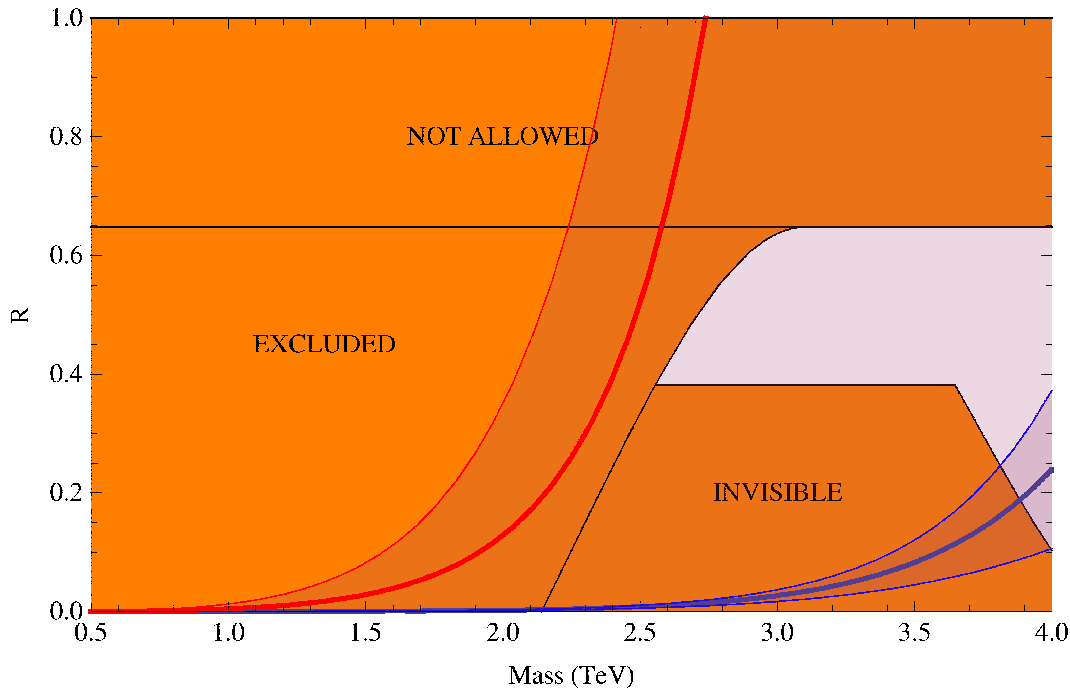}
\caption{\label{reach14}The discovery reach at a machine energy of $\sqrt{s}=$14 TeV and an integrated luminosity of $100~\rm{fb}^{-1}$ for the specific model in the text. Orange (shaded) regions of the plot represent areas of the specific model that are either already excluded,  where Drell-Yan production is suppressed or not accessible by the specific model.   The (thick) blue line represents the minimum coupling ratio $R$ required for at least 10 events to be observed.  (Thin) blue lines follow the contours of an upper and lower $1\s$ PDF uncertainty .  The region between the upper and lower $1\s$ uncertainties is shaded blue.  The (thick) red line represents the minimum coupling ratio $R$ for an asymmetry to be visible at the $3\s$ level (see text for definition).  (Thin) red lines follow the contours of an upper and lower $1\s$ PDF uncertainty (the lower line largely falls below the axis).  The region between the upper and lower $1\s$ uncertainties is shaded red. 
}
\end{figure}

\begin{figure}
\centering
\includegraphics{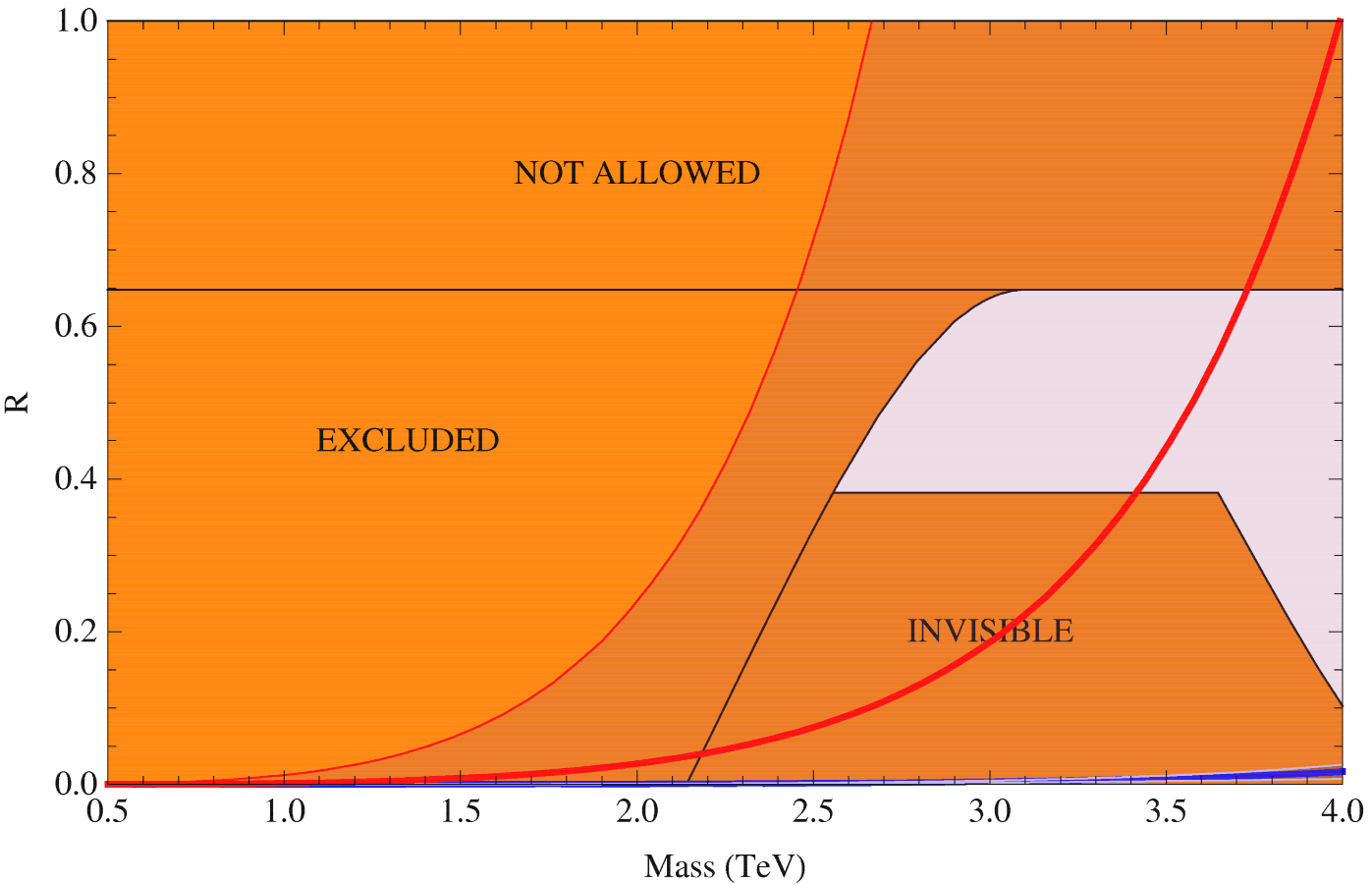}
\caption{\label{reach24}The discovery reach at a machine energy of $\sqrt{s}=$24 TeV and an integrated luminosity of $100~\rm{fb}^{-1}$ for the specific model in the text. Orange (shaded) regions of the plot represent areas of the specific model that are either already excluded,  where Drell-Yan production is suppressed or not accessible by the specific model.   The (thick) blue line represents the minimum coupling ratio $R$ required for at least 10 events to be observed.  (Thin) blue lines follow the contours of an upper and lower $1\s$ PDF uncertainty .  The region between the upper and lower $1\s$ uncertainties is shaded blue.  The (thick) red line represents the minimum coupling ratio $R$ for an asymmetry to be visible at the $3\s$ level (see text for definition).  (Thin) red lines follow the contours of an upper and lower $1\s$ PDF uncertainty (the lower line largely falls below the axis).  The region between the upper and lower $1\s$ uncertainties is shaded red. }
\end{figure}

In this work, several qualitatively different regions of the
$(R,m)$ parameter space are identified.  
Firstly there is a region where the dominant decay mode of the techni-$\r$ particles is into
SM gauge bosons, and not SM fermions. This region is marked as {\it invisible} in figs. ~\ref{reach14} \& ~\ref{reach24} because 
the branching ratio into fermions would strongly suppress the Drell-Yan process of interest here.  
There is then a second region already {\it excluded} by precision measurements.  
Finally the constraint $R\lesssim 0.65$ comes from specific details of the model.

Putting this information into a plot and overlaying the search capabilities of a collider produces Figs.~\ref{reach14} and~\ref{reach24}.
At the very high integrated luminosity of Fig.~\ref{reach14} most of the parameter space will be tested for the existence of new resonances.  
However, there is no reach in terms of asymmetry.  Thus it will be hard to conclude whether there is more than one particle responsible for the peak.  In contrast, with 
the higher collision energy of Fig.~\ref{reach24}, 
a very large fraction of the parameter space presents a testable asymmetry.

Both Figs.~\ref{reach14} \&~\ref{reach24} include a $1\sigma$ PDF uncertainty marked by thin red lines and a shading between.  The lower of these two lines falls below the axis and is barely visible.    
It is clear that the event rate uncertainty is much smaller in the 24 TeV case with the cone barely visible on the plot.  As for the asymmetry, the uncertainty at 14 TeV grows much faster for those masses where the 14 and 24 TeV lines are both able to reach.  In the case of the 14 TeV curve, the uncertainty in coupling is 10 times the coupling estimate by 2 TeV.  In comparison the coupling uncertainty will not reach a factor of 10 until 3.5 TeV in the 24 TeV machine.

For  the representative  point $m=2.6$ TeV and $R=0.4$ we  simulate the asymmetry and see the rapidity cut in action.  For simplicity, we assume that only
$u\bar{u}$ events contribute at the partonic level.  As explained, this is the dominant
subprocess, although as a check simulations including $d\bar{d}$ were run which confirmed expectations or lowering the asymmetry mildly. In this limit, the asymmetry is $A_{FB}\simeq -0.5$.

\begin{figure}
\centering
\includegraphics{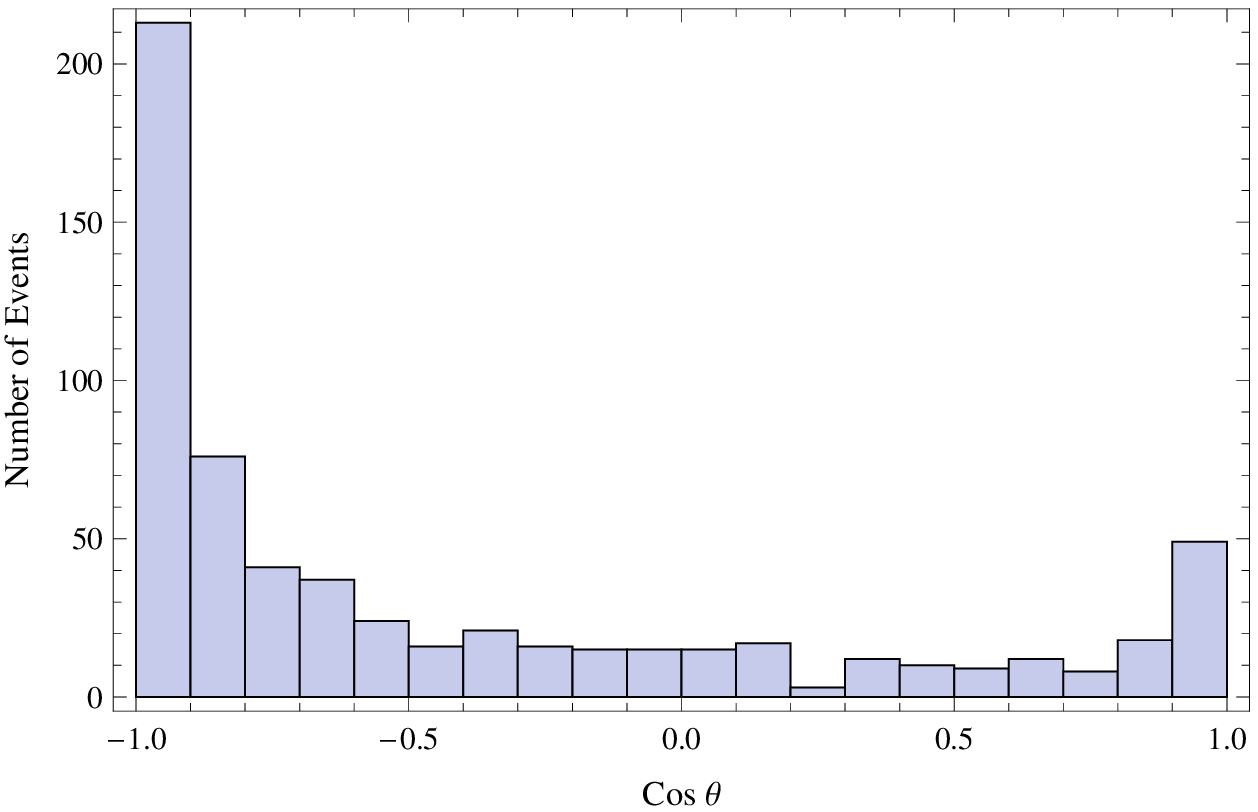}
\caption{\label{24AFBeg} An example simulation of the angular distribution for the predicted number of events at a $\sqrt{s}=$24 TeV machine with $L=100 \,{\rm fb}^{-1}$, for boson mass $m=2.6$ TeV and coupling ratio $R=0.4$.  The simulated data has been restricted to rapidity $Y>Y_c = 1.0$ and the angular distribution in the parton-parton centre of mass frame prediction based on the detector angular distribution is plotted.  The observed asymmetry is $A_{FB}=-0.51 \pm 0.03 ({\rm stat}) \pm 0.1 ({\rm sys})$.  This is the remnant of  initial asymmetry of -0.42 that the data was simulated with.
}
\end{figure}

First let us look at the events with  $\sqrt{s}=24$ TeV.
We apply a cut rapidity $Y>Y_c=1.0$.  In this way the misidentification due to the methodology 
is relatively small ($\simeq  0.1$). Also, the number of events that pass the cut is large enough as to make the statistical element  very small ($\sim 0.03$).
As a result, the simulated set plotted in Fig~\ref{24AFBeg} clearly shows a large asymmetry,
compatible with the true one, and $\sim 4 \sigma$ away from zero.

\begin{figure}
\centering
\includegraphics{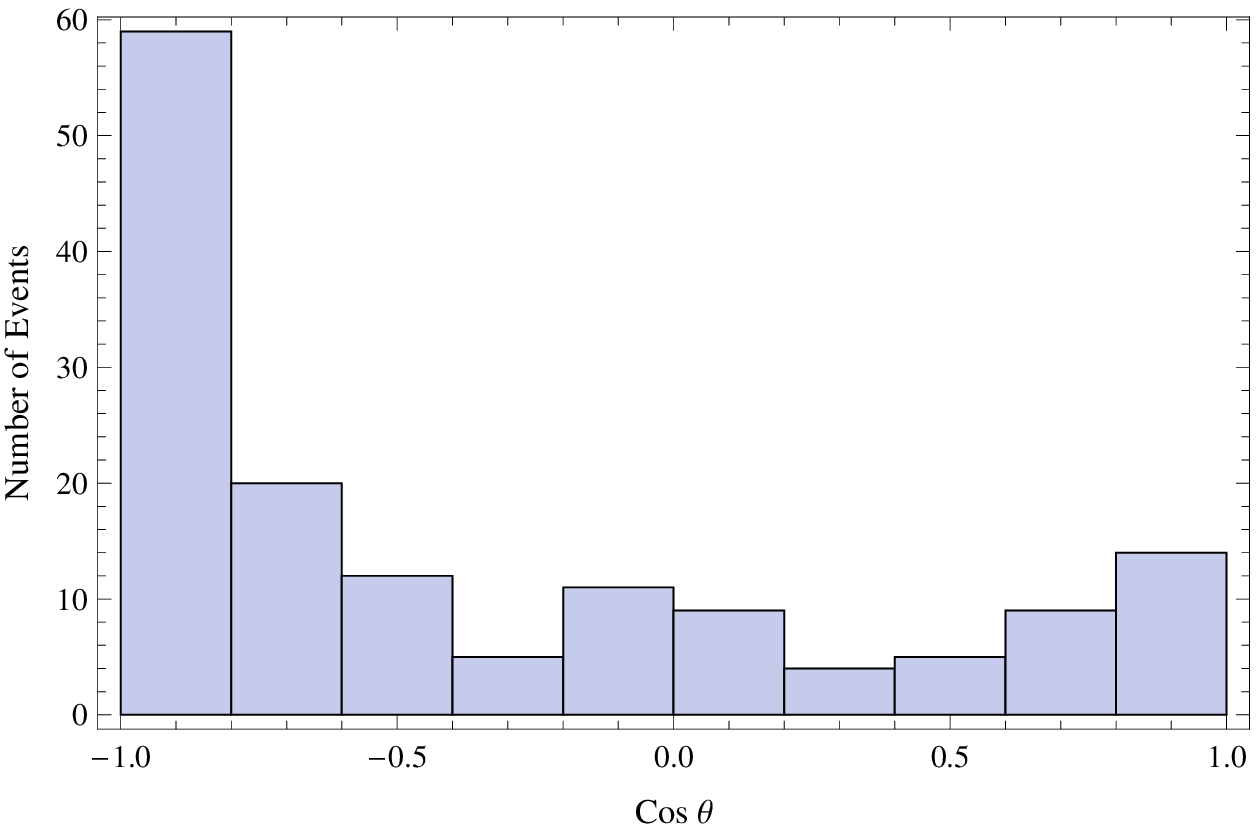}
\caption{\label{14AFBeg}An example simulation of the angular distribution for the predicted number of events at a $\sqrt{s}=$14 TeV machine with $L=100 \,{\rm fb}^{-1}$, for boson mass $m=2.6$ TeV and coupling ratio $R=0.4$.  The simulated data has been restricted to rapidity $Y>Y_c = 0.75$ and the angular distribution in the parton-parton centre of mass frame prediction based on the detector angular distribution is plotted.  The observed asymmetry is $A_{FB}=-0.45 \pm 0.07 ({\rm stat}) \pm 0.14 ({\rm sys})$.  This is the remnant of  initial asymmetry of -0.42 that the data was simulated with.
}
\end{figure}

By comparison, in Fig.~\ref{14AFBeg} we show a set of simulated events with the same 
choices of parameters, but $\sqrt{s}=14$ TeV. In this case we apply a lower cut of $Y>Y_c=0.75$.
As a result, the asymmetry of the simulated data 
is $A_{FB}=-0.45 \pm 0.07 ({\rm stat}) \pm 0.14 ({\rm sys})$, lower than the expected true value, and affected by large uncertainties.

A final comment is in order. In this numerical analysis we did not take into account 
the effect of experimental errors, nor any kinematical cuts that 
might be applied to extract analyzable data.
For example,  when $|\cos\hat{\theta}| \simeq 1$ the experimental extraction of the 
kinematics of the two final leptons might be problematic.
In order to understand how important this effect is,
we extracted the asymmetry from the simulated data by imposing the further constraint 
$|\cos\hat{\theta}|<0.8$. This procedure
effectively reduces the forward-backward asymmetry to
$A_{FB}=-0.37$ ($A_{FB}=-0.28$) in the $24$ ($14$) TeV case.

These two examples show how the method works  in practice.  
However caution must be noted that in performing such simulations there is a degree of sensitivity to the choice of PDF sets.  We illustrate here that the method functions but very different results may be obtained depending on how the large-$\xi$ distributions are modelled,
and ultimately this problem will be resolved only once actual data are collected.

\section{Non-degenerate Couplings}

The region at low rapidity exhibits an interesting feature. 
Due to the differences in the up and down PDF distributions,
the rapidity distributions of these two types of partonic events
have very distinct shapes:
 up-type events peak at finite $Y$, while  down-type peak at $Y=0$.    
 In principle, this might allow to discriminate between different values of the
 couplings $R_{\g^{\prime}}$ and  $R_{Z^{\prime}}$ of the new resonances to the up and down quarks.
 
\begin{figure}
\centering
\includegraphics{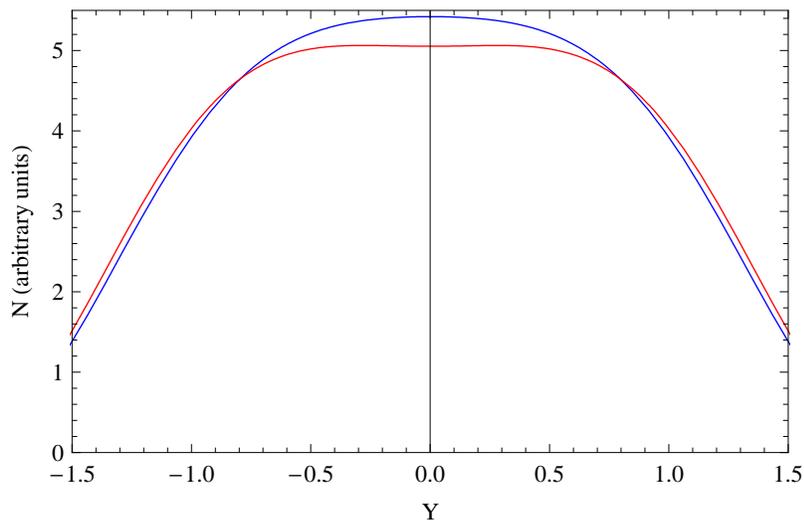}
\caption{\label{Rfind}The expected rapidity distribution of a sample of events in the case that a single new $Z^{\prime}$ (blue) and a single $\g'$ (red) is the only exchange particle present in a new resonance.  The new boson has a mass set at 2.6 TeV and a machine energy of  $\sqrt{s}=24$ TeV is used.  The same number of events is assumed for both.}
\end{figure}

 To investigate the feasibility of such a measurement  we examine the two extreme cases, that $R_{\g'}=0$ versus $R_{Z'}=0$.
We normalize the distributions to the same area, hence  fixing the event number.  
The result is shown in Fig.~\ref{Rfind}.
We see the largest effect is around $Y=0$ where there is a 1 part in 25 difference, or 5\%.  
 The chosen situation is the most extreme, a  light mass
  in terms of what the precision constraints allow and a high energy collider.   
 We can  conclude that the effect of a 
 non-degenerate $R$ will be so small as not to be measurable, even
 assuming perfect detector efficiency and no kinematic cuts.

\section{Conclusions}

We performed a model-independent study of the Drell-Yan processes at the LHC for two new neutral gauge bosons with the quantum numbers of the photon and $Z$ boson.  We assumed the behavior of the new states are analogous to their standard-model cousins, but with couplings to SM fermions rescaled by a common factor of $R<1$.  We also chose to look at the case where the peaks of the two new states were so close as to be indistinguishable  in the invariant mass plots.

We assumed perfect detector performance, omitting experimental errors, 
such as energy resolution, and without including the effect of kinematical cuts.
 We did not include any QCD corrections to the
tree-level partonic calculation of  $pp\rightarrow \mu^{+}\mu^{-}$ and $pp\rightarrow e^{+}e^{-}$. 
In relation to the forward-backward asymmetry, 
we included in the analysis the main sources of theoretical uncertainty: a statistical element
related to the number of expected events, a systematic
 due to the misidentification of the direction of the initial (anti-)quark,
and the uncertainty due to the PDF distributions.

The results can be interpreted as a best case scenario at the LHC, 
limited only by theoretical considerations. We compared  different choices of the integrated
luminosity and center-of-mass energy of the $pp$ collisions.
For illustration purposes, we also referred to one specific model,
in which such mass-degeneracy of the new weakly-coupled  gauge bosons 
is required by imposing the bounds on precision electro-weak parameters.

For a 10 TeV collider, and luminosities in the $2-40$ ${\rm fb}^{-1}$,  bounds on new 
resonances no higher than 3 TeV can be obtained. With a 24 TeV collider the region  
 well above 4 TeV can be explored,  depending on the specific coupling $R$.  
 The uncertainty imparted upon this prediction from the PDFs is quite modest for 
 a high energy collider, while at lower energies, like 10 TeV, the PDF uncertainty
 at large momentum fraction is an important limiting factor.  This is in substantial agreement with the literature.

We studied in detail the forward-backward asymmetry as a method of telling a two particle resonance from a one particle peak.  The collider ability to measure an asymmetry at the $3\s$ level is 
heavily restricted to light masses.  Even at very high luminosities  measuring the asymmetry
 for masses of 3 TeV is not possible at a 14 TeV machine.
Factoring in the PDF uncertainty, it is hard to conceive of any possible asymmetry measurements 
in a 14 TeV (or lower) collider, in relation to the scenarios considered here.  
By contrast, at a 24 TeV machine the PDF uncertainty is markedly lower and the reach far larger.  
At such a machine it may be possible to measure asymmetries and greatly restrict 
currently open parameter space.

In many phenomenological models one would wish to measure precisely the physics of heavy gauge bosons.  
Measuring the forward-backward asymmetry in events with neutral intermediate states  provides a theoretically clean
signal allowing one to discriminate the presence of more than one  boson in a
heavy resonance.   Such a measurement is very difficult in practice from an experimental viewpoint.  On top of the practicalities there are the large PDF uncertainties at large momentum fraction.   Together with 
the scaling of the event rate with $s$, the situation suggests  upgrading the LHC to a higher energy (besides increasing the luminosity) to learn about such new bosons,  if the new bosons are heavy as is suggested by the precision tests.
 
\ack
We would like to thank S.~de~Curtis and M.~Fabbrichesi for useful discussions.  The work of MR is supported by the STFC Doctoral Training Grant ST/F00706X/1, and MP is supported in part  by the Wales Institute of
Mathematical and Computational Sciences and also by the STFC Grant ST/G000506/1.

\end{document}